\documentclass{elsart5p}

\usepackage{natbib}
\usepackage{graphicx}
\usepackage{amssymb}

\begin{document}

\begin{frontmatter}

\title{The Laser Ion Source Trap (LIST) coupled to a gas cell catcher}
\author{T.~Sonoda\corauthref{cor}\thanksref{now1}},
\corauth[cor]{Corresponding author. Tel.: +81-48-467-9428; fax: +81-48-462-4644.}
\ead{tetsu@riken.jp}
\author{T.E.~Cocolios},
\author{J.~Gentens},
\author{M.~Huyse},
\author{O.~Ivanov},
\author{Yu.~Kudryavtsev},
\author{D.~Pauwels},
\author{P.~Van den Bergh},
\author{P.~Van Duppen}

\address{Instituut voor Kern- en Stralingsfysica, K.U.Leuven, Celestijnenlaan 200D, B-3001 Leuven, Belgium}
\thanks[now1]{Present address : RIKEN, 2-1 Hirosawa, Wako, Saitama 351-0198, Japan}

% -----------------------  Abstract ----------------------------------------
\begin{abstract}
The proof of principle of the Laser Ion Source Trap (LIST) coupled to a gas cell catcher system has been demonstrated at the Leuven Isotope Separator On-Line (LISOL). The experiments were carried out by using the modified gas cell-based laser ion source and the SextuPole Ion Guide (SPIG). Element selective resonance laser ionization of neutral atoms was taking place inside the cold jet expanding out of the gas cell catcher. The laser path was oriented in longitudinal as well as transverse geometries with respect to the atoms flow. The enhancement of beam purity and the feasibility for in-source laser spectroscopy were investigated in off-line and on-line conditions. 
\end{abstract}

\end{frontmatter}

% -------------------   Main text ------------------------------------------
% ******************************************************
% ****** Introduction *************************************
% ******************************************************
\section{Introduction}
The laser ion source at the Leuven Isotope Separator On-Line (LISOL) facility provides highly-purified beams of exotic nuclei produced in different types of nuclear reactions \cite{ls1}-\cite{ls7}. The operational principle of the laser ion source is based on element-selective multi-step laser ionization of nuclear reaction products which are thermalized and neutralized inside a high-pressure noble gas. 
The essential part of the laser ion source is the gas cell, which is filled with typically 0.5 bar Ar gas and is placed on the cyclotron beam axis, whereby most ions from the reaction products are neutralized. 
The neutralized atoms are transported by a gas flow towards the exit hole of the cell, where the atoms are re-ionized by laser radiations. The highly-purified beams are thus realized by separation by Z via laser ionization and by A/q at the mass separator. In many cases, however, the mass-separated beam contains in addition to the isotope of interest,  small amounts of isobaric or doubly-charged ion contaminants that survive the neutralization or charge-exchange processes inside the gas cell. For studies of $\beta$-$\gamma$ and $\gamma$-$\gamma$ spectroscopy, such contaminants are unwanted background even if those yields are limited. In order to remove such contaminants, two different approaches are under investigation at LISOL: `the dual-chamber laser ion source' \cite{shd} and the Laser Ion Source Trap `LIST'. The dual-chamber laser ion source is the subject of a separate publication \cite{shd} and only the LIST coupled to a gas catcher is discussed here.

\begin{figure}
\begin{center}
 \includegraphics*[width=\columnwidth]{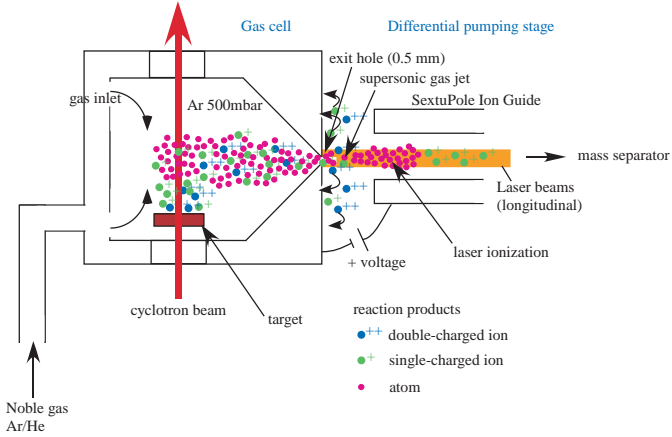}
\end{center}
\caption{The concept of the LIST method coupled to a gas cell catcher.}
\label{fig1}
\end{figure}

The LIST was originally proposed to improve the quality of the ion beam from a hot cavity laser ion source \cite{list1}. This method is also being developed at Jyv\"askyl\"a \cite{list2}\cite{list3}, where the LIST is coupled to the IGISOL gas cell catcher \cite{ig}. The concept of the LIST method coupled to a gas cell catcher is shown in Fig.~\ref{fig1}. The reaction products are thermalized and stopped in the buffer gas, subsequently neutralized and finally flushed out of the gas cell in a supersonic gas jet. The resonance ionization from the laser beams takes place in the gas jet leaving the gas cell and the photo-ions are captured in the RF-field of the SextuPole Ion Guide (SPIG)\cite{sp1}-\cite{sp4} located immediately after the gas cell. In order to suppress unwanted ions, a positive DC voltage is applied on the SPIG rods to prevent the surviving ions that escape the gas cell from entering the SPIG. Only the ions that are resonantly ionized by the lasers close to the entrance of or inside the SPIG are sent through the mass separator. The purity of the beam can be further improved by applying a time gate after each laser pulse. In this way,  unwanted ions can be further suppressed yielding extreme purity of the final beam.

The beam-quality improvements reached with the LIST open new possibilities such as in-source laser spectroscopy. For precise laser-spectroscopic studies,  the width of the measured resonant spectral line should be as narrow as possible. The experimental width is a convolution of many effects adding to the intrinsic laser linewidth, including mainly the pressure broadening from collisions with the buffer gas, the power broadening from the lasers and the Doppler broadening due to the atomic velocity distribution. In the LIST case, the isolated atomic beams are obtained by supersonic adiabatic expansion in vacuum, which reduces the Doppler broadening substantially. Additionally, the gas density in the ionization region is too low to significantly contribute to the pressure broadening of the spectral line. While in some cases elements with large hyperfine structure or large isotope shift can be studied inside the gas cell \cite{backe}\cite{Yean}, the LIST condition creates a suitable environment for laser spectroscopy on a wider range of nuclei, complementarily to standard ISOL systems making use of solid or liquid target/catcher systems. In those cases, the release properties can seriously reduce the efficiency for certain elements or for short-living nuclei. Here, the time restriction from the decay losses is only the evacuation time of the gas cell. Short-lived isotopes, with half-lives down to 100 ms, are suitable candidates for laser spectroscopy.

In the present work, we studied the LIST performances in off-line and on-line conditions. The laser beams are sent either by a longitudinal or a transverse path with respect to the gas jet outside the cell. In the off-line conditions, laser beams ionize stable Co, Ni or Cu atoms evaporated from a filament located inside the cell. The suppression effect of unwanted ions is shown by monitoring the time profile on the arrival of mass-separated ions while applying different repeller voltages. Frequency scans of the first step laser for stable $^{58}$Ni and $^{63}$Cu in the gas cell and in the jet have been performed to compare the resonant linewidth in the different conditions. The evolution of the pressure broadening and the pressure shift in argon as a function of the argon pressure in the cell for the resonant 232.003 nm nickel line and the resonant 244.164 nm copper line were evaluated. As a demonstration, the isotope shifts of $^{58,60,62,64}$Ni have been measured in the jet. In on-line condition, neutron-deficient Rh isotopes produced in fusion-evaporation reaction were successfully ionized in the LIST.
% ******************************************************************
% ************** Experimental setup ***********************************
% ****************************************************************** 
\section{Experimental set-up}

\begin{figure}
\begin{center}
 \includegraphics*[width=\columnwidth]{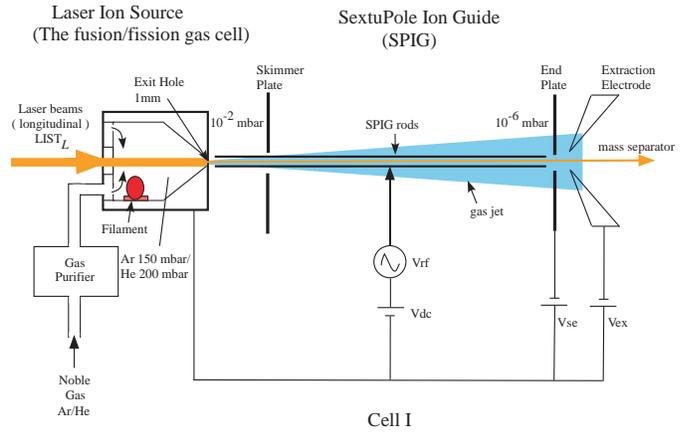}
\end{center}
\caption{A top view of the single-chamber gas cell together with the SPIG in the LIST experiment.}
\label{fig2}
\end{figure}

\begin{figure}
\begin{center}
 \includegraphics*[width=\columnwidth]{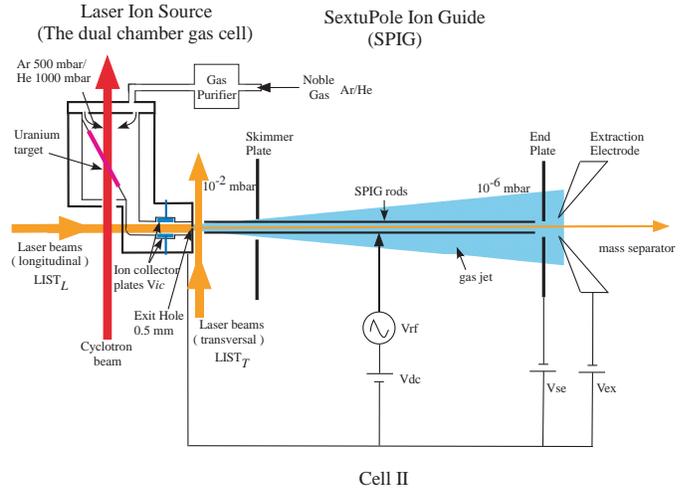}
\end{center}
\caption{A top view of the dual-chamber gas cell \cite{shd} together with the SPIG in the LIST experiment.}
\label{fig3}
\end{figure}

Two different gas cells were used for the LIST experiment. The experimental set-ups are shown in Fig.~\ref{fig2} and \ref{fig3} according to the cell type. The first cell, which is a single-chamber gas cell, is shown in Fig.~\ref{fig2}. It was originally used for heavy ion-induced fusion-evaporation and proton-induced fission reactions in on-line LISOL experiments \cite{ls6}. In the present test, it was used only for off-line measurements. The diameter of the exit hole was 1 mm and the gas pressure in the cell was fixed at 150 mbar of argon and 200 mbar of helium respectively.

The second cell is the dual-chamber gas cell, shown in Fig.~\ref{fig3}. A detailed description of this gas cell can be found in \cite{shd}. Its main feature is the separation in two volumes: one for thermalizing the reaction products and the other for laser re-ionization. Due to a lower charge density in the re-ionization volume, electrical fields inside the gas cell can be applied using a potential $V_{ic}$ on the ion collector plates (Fig.~\ref{fig3}). The diameter of the exit hole was 0.5 mm and the pressure of the cell can be increased up to 500 mbar for argon and 1000 mbar for helium. The maximum pressure in both cells thus depends on the size of the exit hole and the pumping capacity outside the cell. The dual-chamber gas cell was used for both off-line and on-line experiments. In the on-line test, an accelerated beam from the cyclotron impinged on a target which was tilted by 35 degrees with respect to the beam direction. The reaction products recoiling out from the target are thermalized inside the stopping volume and then move to the re-ionization volume by a gas flow. 

The SPIG is located at the differential pumping region  in front of the exit hole of the cell. This technique was originally proposed for the ion transportation from the low- to high-vacuum regions while maintaining the beam quality \cite{sp1}. The distance between the gas cell and the SPIG is adjustable. The six rods of the SPIG have a diameter of 1.5 mm, the length of the rods is 126 mm and the diameter of the inner circle of the ion guide is 3 mm. The voltage configuration consists of 3 parameters: SPIG $V_{rf}$ (radio-frequency, typically 300 Vpp, 4.7 MHz), SPIG $V_{dc}$ (superimposed to SPIG $V_{rf}$ before being applied to the SPIG rods), and SPIG-end $V_{se}$.  In order to see the laser ions which are produced inside the SPIG, a positive potential was applied to SPIG $V_{dc}$ repelling unwanted ions coming from the gas cell. The SPIG-end $V_{se}$ was given a negative or zero voltage. The acceleration voltage on the isotope separator is typically set at 40 kV.

The optical system has been thoroughly described in Ref.~\cite{ls2}. It consists of two tunable dye lasers pumped by two XeCl excimer lasers. The maximum laser pulse repetition rate is 200 Hz. Two-color, two-step schemes are used to ionize atoms through auto-ionizing states. Two laser paths to the LIST were used either in the longitudinal (LIST$_L$) or in the transverse (LIST$_T$) direction with respect to the atom beam. In the LIST$_L$, the lasers are introduced from the backside of the cell and pass through the exit hole and the SPIG. In the LIST$_T$, the lasers come across the gas jet between the exit hole of the cell and the SPIG, perpendicular to the gas jet. The single-chamber gas cell was only tested with the LIST$_L$, while the dual-chamber gas cell was studied with both geometries. For atomic-spectroscopy studies, the intrinsic bandwidth of the first-step laser has been minimized to 1.6 GHz with etalon, starting from 4.5 GHz in the second harmonic without etalon. 

A fraction of the laser beams is deflected into a reference cell, where an atomic beam of the investigated element is produced from a resistively-heated crucible. The pressure in the reference cell is $10^{-6}$ mbar. The laser beams ionize the atoms in a crossed-beam geometry and the obtained ions are accelerated towards a secondary electron multiplier. 
This set-up is used to perform laser spectroscopy in vacuum. 
Furthermore, the wavelength of the first step transition is consistently monitored by a Lambdameter LM-007.

In the present study, there are a number of limitations. First, when the lasers are sent through the gas cell (LIST$_L$), the ionization region where the laser beams and the jet atoms overlap is restricted due to a limited laser-spot size determined by the diameter of the gas cell exit hole. This can be solved by sending the lasers from the other end through the isotope separator and the acceleration electrodes \cite{list2}; it is however not possible with the current setup, as the dipole magnet has no window. The size of the exit hole being 1 mm or 0.5 mm in diameter, the ionization region will also be a cylinder of that dimension. This value is more than 6 times smaller than the original laser spot size and reduces greatly the overlap of the laser beam with the plume of atoms and thus the ionization efficiency. This limitation is not fully avoided even if sending the lasers in transverse geometry, where also the size of the expanding jet is larger than the laser cross section.

Moreover, our present laser system is not ideal for the LIST as the maximum repetition rate of the pulsed lasers is 200 Hz. If the overlapped length in laser photons and jet atoms, in the longitudinal mode, is 50 mm and the jet velocity is 500 m/s, then 10 kHz repetition rate is required for at least one encounter between the laser photons and the atoms.  Therefore the present setup reaches only up to 1/50 of the LIST capability. This is even worse in the transverse mode. Additionally, the laser bandwidth after frequency doubling is 1.6 GHz, which is considered wide for laser spectroscopy. Due to those limitations, the present work represents a feasibility study and not yet a report on a full-fetched facility. However, since this is expected to be a linear behavior, one can scale the measured efficiencies with this reduction factor of 50 in duty cycle to estimate the performance of the LIST mode.  

% ******************************************************************
% ************** Results and discussion ***********************************
% ****************************************************************** 
\section{Results and discussion}

The different approaches used in this work are detailed in Table \ref{tbl:conditions}. The section is then divided according to the type of laser path used in the LIST: longitudinal (LIST$_L$) or transverse (LIST$_T$). 

\begin{table}
 \caption{Experimental conditions for the gas cell, the longitudinal LIST$_L$ and the transverse LIST$_T$. Two voltage configurations, SPIG V$_{dc}$ and IC(Ion Collector) V$_{ic}$, are adjusted for the different conditions.\label{tbl:conditions}  }
 \begin{center}
 \begin{tabular}{cccc}
  \hline
  & SPIG $V_{dc}$ (V) & IC $V_{ic}$ (V) & Laser path\\
  \hline
  Gas cell & -210 & 0 & longitudinal\\
  LIST$_L$ & +46 & 0 & longitudinal\\
  LIST$_T$ & 0 & $\pm40$ & transverse\\
  \hline
 \end{tabular}
 \end{center}
\end{table}

\subsection{LIST using longitudinal laser ionization - LIST$_L$}

\subsubsection{Suppression of unwanted ions with the repeller voltage}

\begin{figure}
\begin{center}
 \includegraphics*[width=\columnwidth]{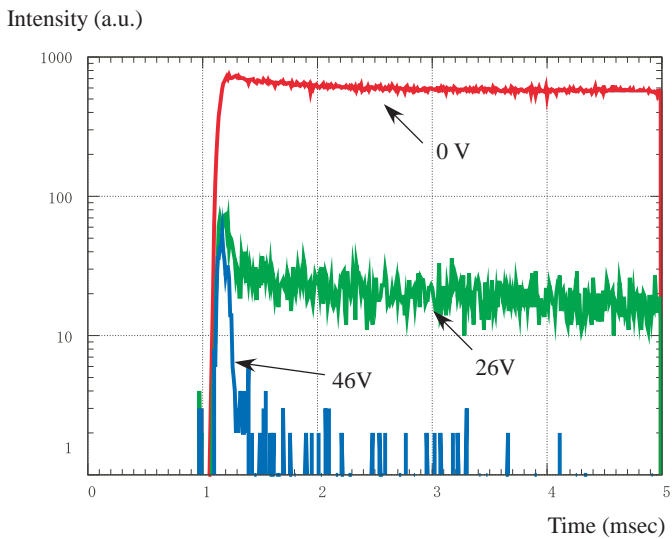}
\end{center}
\caption{The time profile of Cobalt ions ($^{59}$Co$^{1+}$) transported via the SPIG and mass-separated (colour on-line). The laser pulse was fired at t=1 ms. Different SPIG $V_{dc}$ voltages were used as indicated in the figure. The voltage polarity was applied positively to suppress ions from the gas cell. The delay between the laser pulse and the start of an ion pulse is 100 $\mu$sec.}
\label{fig4}
\end{figure}

In order to suppress unwanted ions coming from inside the cell, a positive potential was applied to SPIG $V_{dc}$ with respect to the cell. Fig.~\ref{fig4} shows the result of time-profile measurements of stable cobalt ions ($^{59}$Co$^{1+}$) with different SPIG $V_{dc}$ voltages. The gas cell 'Cell I', shown in Fig.~\ref{fig2}, was used. The mass-separated ions were counted by the Secondary Electron Multiplier (SEM) located one meter downstream from the focal plane at the end of the mass separator. The laser pulse was fired at t=1 ms longitudinally through the gas cell filled with Ar at pressure 150 mbar and passed through the 1 mm exit hole into the SPIG. The laser-repetition rate was 1 Hz, avoiding effects from previous pulses. The SPIG-end $V_{se}$ was fixed at $-40V$, the distance between the SPIG and the exit of the cell was 2 mm. 

In Fig. \ref{fig4}, when the repeller voltage is 0 V, the majority of the signal is made by ions ionized inside the gas cell from which they are continuously evacuated. When applying a positive SPIG $V_{dc}$ voltage, most photo-ions from the gas cell are prevented from entering the SPIG. However even at 26 V, ions from the gas cell are still entering the SPIG due to the continuous collisions with the jet atoms. By increasing the voltage further, the ions coming from inside the cell are finally suppressed from entering the SPIG. Consequently, the remaining signal in the time profile comes from ions which are ionized only outside the gas cell and captured by the SPIG. Some counts also appear at t = 1 ms in Fig.~\ref{fig4}, produced by scattered laser photons entering the SEM. This contributes a few counts in total SEM signals. The background of SEM without any lasers was nearly zero.
The delay between the laser pulse and the start of an ion pulse was 100 $\mu$s, from which $\sim20$ $\mu$s correspond to the time of flight of the ions through the separator; the remaining 80 $\mu$s correspond to the transport time through the SPIG. The signals left over at longer time with the highest voltage ($V_{dc}$= 46 V) are ions that are delayed in the SPIG. The width of the remaining peak was about 88 $\mu$sec in Full Width at Half Maximum (FWHM).

Similar results were also observed with He as the buffer gas at a pressure 200 mbar. The suppression voltage needed to observe the ion signal from the LIST mode was about 20 V. 

\subsubsection{Wavelength scans and the velocity evaluation of the gas jet}

\begin{figure}
\begin{center}
 \includegraphics*[width=\columnwidth]{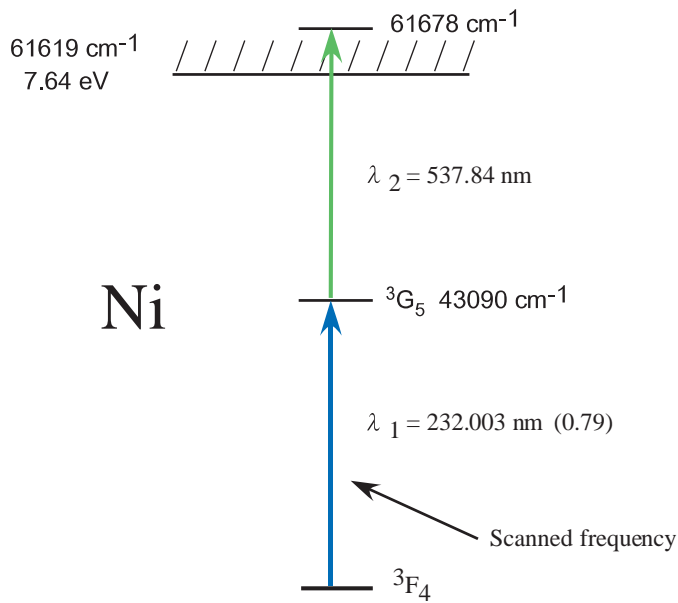}
\end{center}
\caption{Partial atomic level diagram of Ni. Next to the wavelength, the log-ft(transition strength) is listed \cite{strength}.}
\label{fig5}
\end{figure}

One of the interests in the LIST mode is to study the feasibility of laser spectroscopy inside the gas jet for exotic nuclei.
Therefore the resonant linewidth of a specific element has to be evaluated in the gas jet and  compared to other conditions.
The laser ion source at LISOL allowed for direct comparison of the resonant linewidth for stable isotopes in the following three circumstances:
1) inside the reference cell (vacuum, 10$^{-6}$ mbar), 2) inside the gas cell (He or Ar with a few hundreds mbar), 3) inside the gas jet (LIST$_L$/LIST$_T$).
Frequency scans of the first step laser for stable $^{58}$Ni have been performed and the resonant linewidth at those different locations is extracted.
The partial atomic level scheme of Ni is given in Fig.~\ref{fig5}. An efficient ionization path is used, starting from the $^3$F$_4$ ground state via a transition at $\lambda_{1}=232.003$ nm to the $^3$G$_5$ intermediate level at 43090 cm$^{-1}$, followed by a transition at $\lambda_{2}=537.84$ nm to an auto-ionizing state. 

\begin{figure}
\begin{center}
 \includegraphics*[width=\columnwidth]{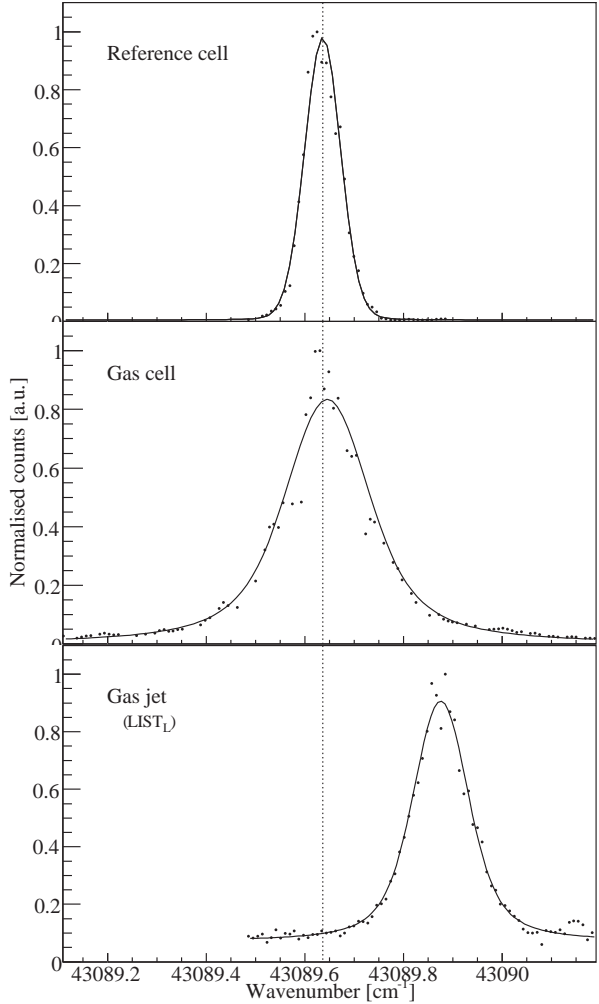}
\end{center}
\caption{Scan of the first step transition of Ni in three different locations: in the reference cell (top), in the gas cell (middle), in the LIST$_L$ (bottom), using 200 mbar helium as the buffer gas. A Doppler shift of 7.2 GHz is observed when ionizing in the jet, which corresponds to an atom velocity of 1663m/s. The solid line represents the best fit of a Voigt profile through the data points.}
\label{fig6}
\end{figure}

Fig.~\ref{fig6} shows the resonance of the first step transition under the three different conditions. In this measurement, the Cell I configuration (Fig.~\ref{fig2}) was used with 200 mbar of helium as the buffer gas. The SPIG $V_{dc}$ voltage was kept at +46V in the LIST mode. The result clearly shows a displacement of the resonance centroid acquired in the jet from that acquired in the reference cell or the gas cell. This displacement is caused by the Doppler shift of the moving atoms in the jet while they are ionized inside the SPIG. From this measurement, the jet velocity was deduced to be $\approx$ 1663 m/s using the displacement $(\nu'-\nu)=+7.2$ GHz, where $\nu'$ is the resonance frequency of atoms in the jet and $\nu$ is the resonance frequency for atoms in the reference cell.

Table \ref{tbl:table5} shows the values of the centroid and the Full Width at Half Maximum (FWHM) of the resonance peak in those different locations. There are four components which should be convoluted in this resonance line. The first component is the intrinsic band width of the laser ($\Delta_{laser}\approx1.6$ GHz Gaussian FWHM in this setup). The second component is a laser power broadening $\Gamma_{power}$ (Lorentzian FWHM). This was a very important source of broadening; the laser power was therefore adjusted to a value as low as possible in all three cases while still allowing sufficient ionization to perform the measurement. The two other components are the Doppler broadening $\Delta_{Doppler}$ (Gaussian FWHM) from the atom velocity distribution and the pressure broadening $\Gamma_{pressure}$ (Lorentzian FWHM) from the surrounding gas. This last component grows linearly with the pressure as
\begin{equation}\label{eqn:pressure}
\Gamma_{pressure} = \textrm{constant}\cdot\textrm{Pressure},
\end{equation}
where the constant depends on the atomic transition of interest. The overall resonance line is therefore a Voigt profile where the Gaussian and the Lorentzian contributions are given by
\begin{eqnarray}
 \Delta & = & \sqrt{\Delta^{2}_{laser} + \Delta^{2}_{Doppler}}\label{eqn:delta}\quad\textrm{and}\\
 \Gamma & = & \Gamma_{power} + \Gamma_{pressure}.\label{eqn:gamma}
\end{eqnarray}
 The total FWHM is then given empirically by \cite{Oli77}
\begin{equation}\label{eqn:FWHM}
\textrm{FWHM} = 0.5346\cdot\Gamma + \sqrt{0.2166\cdot\Gamma^{2} + \Delta^{2}}.
\end{equation}

\begin{table}
\caption{Comparison of the linewidth of Ni in different locations using He 200 mbar as the buffer gas based on the data in Fig.\ref{fig6}.}
\begin{center}
\label{tbl:table5}
\begin{tabular}{lr@{.}lr@{.}l@{/}r@{.}l}
\hline
Ionization place   & \multicolumn{2}{c}{Centroid (cm$^{-1}$)} & \multicolumn{4}{c}{FWHM (cm$^{-1}$/GHz)} \\
\hline \hline
Reference cell     & 43089&636      & 0&101(5) &3&03(15)\\ % run 17
Gas cell           & 43089&646(15)  & 0&211(33)&6&33(99)\\ % run 51
LIST$_L$           & 43089&875(14)  & 0&135(6) &4&05(18)\\ % run 37
\hline
\end{tabular}
\end{center}
\end{table}

In the wavelength scan for the evaporated atoms in the reference cell, the pressure broadening is negligible as high vacuum is reached inside the reference cell: $\Gamma\approx\Gamma_{power}$. This contrasts strongly with the gas cell where the pressure broadening is typically dominant contribution for the width of a resonance peak, that depends on the type of element, electron transition and the amount of gas pressure. This pressure effect is evaluated by the resonance linewidth when the pressure is systematically changed. Additional remark in the case of the reference cell, the ionization is performed in a crossed-beam geometry, thus probing the atomic beam in a direction where the velocity is  perpendicular; the Doppler broadening is thus negligible: $\Delta\approx\Delta_{laser}$.  As for the Doppler broadening in the case of the gas cell, the velocity of the atoms in the gas is subjected to a Maxwell-Boltzmann distribution. The velocity range depends on the mass, with a wider distribution for lighter masses. In the case of Ni, a simple calculation yields a velocity range of FWHM$\approx300$ m/s in 200 mbar He as a buffer gas. This value broadens the peak by $\Delta_{Doppler}\approx2$ GHz.

In the case of ionization in the SPIG, the jet conditions are the most important. As the pressure in the entrance of the SPIG is already low, the pressure broadening is reduced substantially.  The main contribution to the FWHM is therefore the Doppler broadening in addition to an intrinsic band width and a power broadening of the laser. 

\subsubsection{Pressure broadening of the nickel resonance line in argon}

\begin{figure}
\begin{center}
 \includegraphics*[width=\columnwidth]{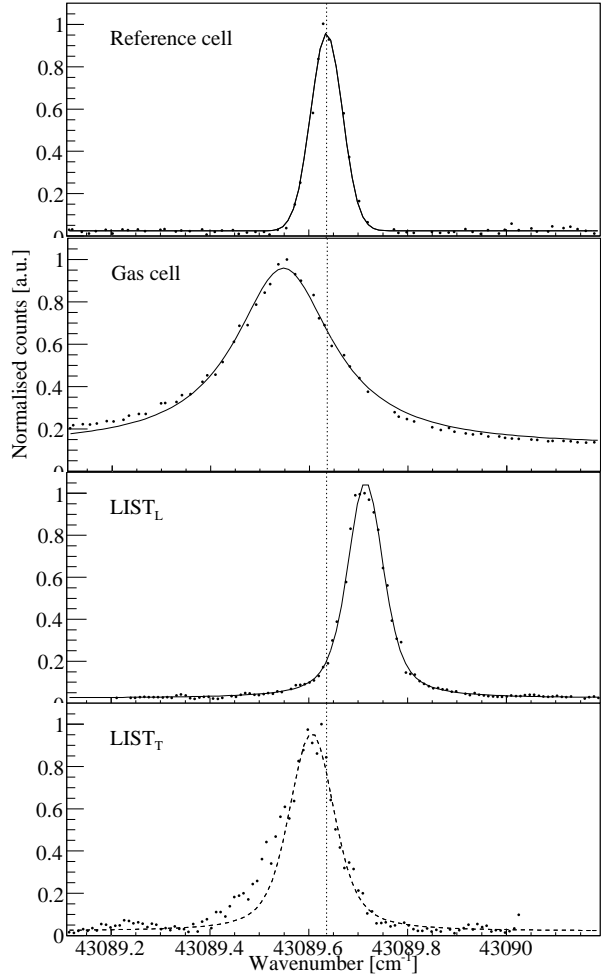}
\end{center}
\caption{The resonant linewidths of the first step transition of Ni in three different locations: in the reference cell (top), in the gas cell (second from the top), and in the gas jet with the lasers in the longitudinal LIST$_{L}$ (second from the bottom), and with the lasers in the transverse LIST$_{T}$ (bottom). The gas cell was filled with 500 mbar of argon as the buffer gas. The solid line is the best fit of a Voigt profile through the data point. The dashed line is the best fit through the high-frequency half of the asymmetric resonance in the LIST$_{T}$; the asymmetry is due to the high pressure gradient in the region close to the exit nozzle.}
\label{fig7}
\end{figure}

In the gas cell with an exit hole diameter of 1 mm, the maximum argon pressure of 150 mbar is limited by the pumping capacity of the system. With an exit hole of 0.5 mm, the gas pressure can be increased up to 500 mbar. At this pressure, the broadening and the shift of the nickel resonant line are large enough to be measured with the existing laser bandwidth. Fig.~\ref{fig7} shows the wavelength scans of the first step laser in the three different locations. The deduced FWHM values are given in Table \ref{tbl:table8} in the same form as in Table \ref{tbl:table5}. The width of the resonance in the reference cell and in the jet (LIST$_L$) are the same around $\sim$2 GHz. The signal from the gas cell filled with 500 mbar argon is, however, much broader ($\sim$6 GHz) and red-shifted by 2.5 GHz relative to the resonance in the reference cell. Similarly to the case where He was used, the jet velocity was deduced by the displacement of the SPIG resonance peak and resulted in $\approx$ 550 m/s.
 
\begin{figure}
 \begin{center}
  \includegraphics*[width=\columnwidth]{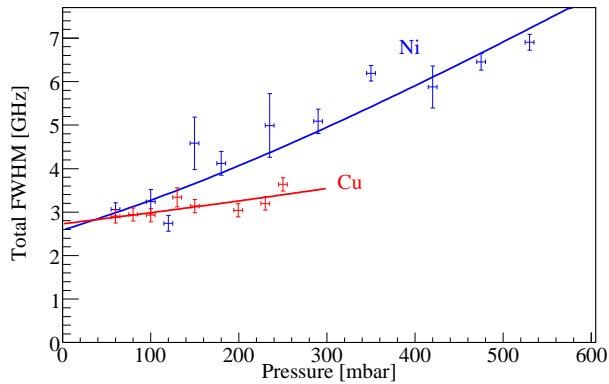}
 \end{center}
  \caption{The evolution of the pressure broadening in argon as a function of the argon pressure in the cell for the resonant 232.003 nm nickel line and the resonant 244.164 nm copper line. Some data points are the average of several measurements. The solid lines are the best fits according to equations \ref{eqn:pressure}-\ref{eqn:FWHM}.}
 \label{fig:fig8}
\end{figure}

\begin{figure}
 \begin{center}
  \includegraphics*[width=\columnwidth]{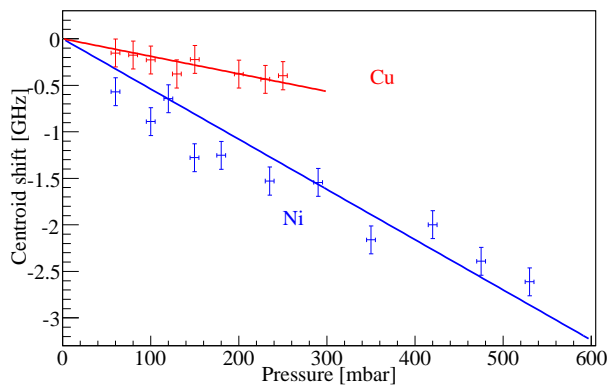}
 \end{center}
  \caption{The evolution of the pressure shift in argon as a function of the argon pressure in the cell for the resonant 232.003 nm nickel line and the resonant 244.164 nm copper line. Some data points are the average of several measurements. The solid lines are linear best fits going through the origin.}
 \label{fig:fig9}
\end{figure}

Performing similar comparisons at different argon pressures gives the evolution of both the pressure broadening and the pressure shift. The results are shown in Fig.~\ref{fig:fig8} and \ref{fig:fig9} for the resonant 232.003 nm nickel line and the resonant 244.164 nm copper line. A pressure broadening of 11.3(6) MHz per mbar and a pressure shift of $-5.5(3)$ MHz per mbar can be extracted for nickel; in the case of copper, a pressure broadening of 5.4 MHz per mbar is found and a pressure shift of -1.9(1) MHz per mbar. The difference between nickel and copper highlights the importance of the electronic transition studied.

\begin{table}
\caption{Comparison of the linewidth of Ni in different locations using 500 mbar of Ar as the buffer gas based on the data in Fig.\ref{fig7}. Most of the uncertainty comes from systematic effects of the laser power fluctuations and laser modes; it has been estimated to 0.005 cm$^{-1}$ based on the fluctuations observed on the spectra of Fig.~\ref{fig7}.}
\begin{center}
\label{tbl:table8}
\begin{tabular}{lr@{.}lr@{.}l@{/}r@{.}l}
\hline
Ionization place   & \multicolumn{2}{c}{Centroid (cm$^{-1}$)} & \multicolumn{4}{c}{FWHM (cm$^{-1}$/GHz)} \\
\hline \hline
Reference cell     & 43089&636     & 0&064(5) &1&92(15)\\  % run December #3
Gas cell           & 43089&551(5)  & 0&215(5) &6&45(15)\\  % run December #3
LIST$_L$           & 43089&715(5)  & 0&087(7) &2&61(21)\\  % run March #12
LIST$_T$           & 43089&606(33) & 0&108(15)&3&24(45)\\  % run June #4
\hline
\end{tabular}
\end{center}
\end{table} 

\subsection{LIST using transverse laser ionization - LIST$_T$}

\subsubsection{Suppression of unwanted ions with the collector plates (ion collector)}

The repeller voltage to suppress unwanted ions used in the longitudinal geometry cannot be used in the transverse geometry as the ions produced between the gas cell and the SPIG would all be repelled. Instead, a voltage ($V_{ic}$) is applied to the ion collector inside the laser ionization chamber, as described in \cite{shd}.

This method can only be used with the dual-chamber gas cell pictured in Fig.~\ref{fig3}. A voltage difference from $-40$ V to $+40$ V is applied across the plates to collect the ions surviving the neutralization processes in the gas catcher and only an atom beam exits the cell. The ions are then produced between the gas cell exit and the SPIG, placed at a distance of $3$ mm, and at the entrance of the SPIG, since the laser spot size is 5 mm.

The performance of the ion collector depends on several parameters such as the collection efficiency of the ions and the ion production rate is discussed in \cite{shd}.

\subsubsection{Overlap between the laser and atom beams}

Compared to the longitudinal mode inside the gas cell and based on a laser repetition rate of 200 Hz, a laser spot size of 5 mm and a supersonic velocity in the Ar gas jet of 560 m/s, a reduction factor of 560 in duty factor is expected in this transverse mode. This assumes that both the laser excitation and ionization steps are saturated. Experimentally, a reduction factor of $300(10)$ has been measured with $^{58}$Ni$^+$ ions from a filament, probably due to a larger laser-spot size than in the previous estimate.

\begin{figure}
\begin{center}
 \includegraphics[width=\columnwidth]{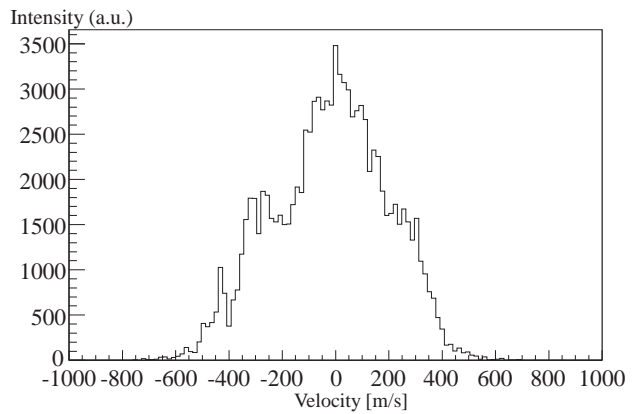}
 \caption{Simulated velocity distribution in the direction perpendicular to the atom jet in the area covered by the lasers between the gas catcher and the SPIG. The buffer gas in the gas catcher is Ar at 500 mbar. This distribution contributes to the total broadening of the optical resonance by $750$ MHz.}
 \label{fig:fig10}
\end{center}
\end{figure}

Another important parameter concerning the overlap of the two beams in this geometry is the velocity distribution of the atoms. With the increased distance between the gas cell exit and the SPIG, the atom beam diverges strongly; a simulated velocity distribution in the transverse direction is shown in Fig.~\ref{fig:fig10}. Such a distribution translates into a Gaussian profile in the optical resonance with a FWHM of 1 to 2 GHz. Convoluted with the laser lineshape, this contributes to the total broadening of the optical resonance; in the case of a gas catcher filled with $500$ mbar of Ar, the estimated increase in broadening is $750$ MHz.

\subsubsection{Wavelength scans and the environmental conditions}

The optical resonance using the LIST$_T$ is shown at the bottom of Fig.~\ref{fig7}. Its properties appear in Table \ref{tbl:table8}.

The broader FWHM in the LIST$_T$ can be partially explained by the transverse velocity distribution previously discussed. However, some effects from the gas pressure could still play a role as there could be a pressure gradient close to the exit nozzle. This pressure gradient is also responsible for the asymmetry and the shift of the resonance in the LIST$_T$.

\subsubsection{On-line measurement}

\begin{figure}
\begin{center}
 \includegraphics*[width=\columnwidth]{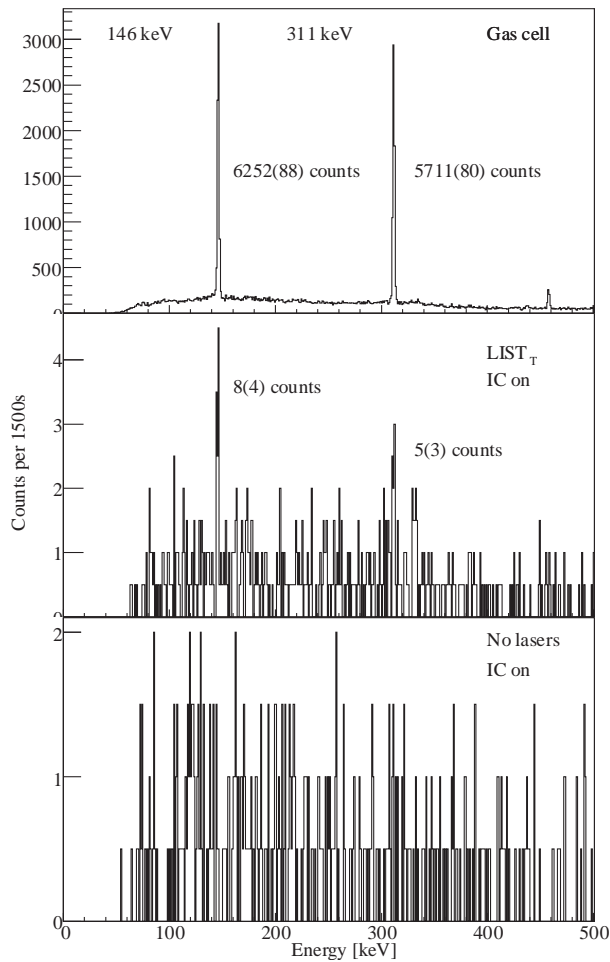}
\end{center}
\caption{$\gamma$ spectra in the decay of $^{94}$Rh from $0$ to $500$ keV. From top to bottom: longitudinal ionization inside the gas catcher; ionization in the LIST$_T$; background in the LIST mode without laser ionization. The spectra are normalized to a measurement time of 1500s.}
\label{fig:fig11}
\end{figure}

The LIST$_T$ was used on-line with radioactive neutron-deficient $^{94}$Rh isotopes produced in the $^{58}$Ni($^{40}$Ar,1p3n)$^{94}$Rh reaction. In this heavy-ion reaction and using longitudinal ionization in the gas cell, the 70.6 s ($4^+$) low-spin ground state is produced with 900(100) ions per $\mu$C while the 25.8 s (8$^+$) high-spin isomer, more favored, is produced with 6000(100) ions per $\mu$C. The $\gamma$ spectra in the different conditions are shown in Fig.~\ref{fig:fig11}.

By comparing longitudinal ionization inside the gas catcher to transverse ionization in the LIST$_T$, a reduction factor in efficiency of at least 700 is extracted from the $\beta$-decay rates. This is more than in the case of stable $^{58}$Ni although it is of a similar order of magnitude. This could be due to the difference in the laser effective spot size between those elements.

On the other hand, the difference between the LIST mode with and without laser ionization is striking. The peaks are not visible anymore once the lasers are blocked while they are still clear with the laser ionization. Although the limited statistics only allows the extraction of a lower limit on the selectivity of 4, the real value that can be expected is much larger.

In case of the neutron-rich isotopes produced in the proton-induced fission of $^{238}$U, another source of contamination is present through the deposits of neutral radioactive isotopes on the RF structure of the SPIG  \cite{shd}. The subsequent $\beta^-$ decay of these neutron-rich nuclei leaves the daughter nuclei in an ionic state, yielding possibly in the capture by the pseudo-potential of the SPIG. This could be a limit of applicability of the LIST concept for the neutron-rich isotopes.

\subsection{Laser spectroscopy in and around a gas catcher}

\begin{figure}
\begin{center}
 \includegraphics*[width=\columnwidth]{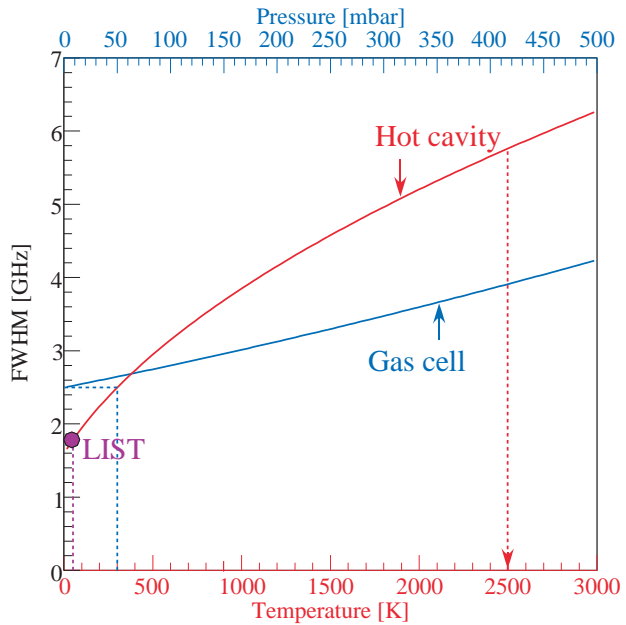}
\end{center}
\caption{Simulated resonance linewidth of the copper transition at 244.164 nm for in-source laser spectroscopy with a hot cavity as a function of the atomizer temperature and for in-gas-cell laser spectroscopy at room temperature (with $\Delta_{Doppler}$(300 K)) as a function of the gas cell pressure for the typical working range of LISOL (colour on-line). A typical inherent laser bandwidth of 1.6 GHz is assumed in both cases. The typical running temperature for the hot cavity (ISOLDE, 2500 K \cite{CuHFS1}-\cite{CuHFS2}) and for the gas cell are shown with dashed lines. The LIST operating mode is also shown (low temperature and low pressure) and the resolution is dominated by the total laser bandwidth.}
\label{fig:fig12}
\end{figure}

Laser spectroscopy in ion-sources has been already performed in both hot cavity ion sources \cite{Gatchina} and gas catchers \cite{backe}. The resolving power of each technique can be compared by analyzing the resonance linewidth in their respective type of ion source. In the case of in-source laser spectroscopy with a hot cavity \cite{CuHFS1}\cite{PbCR}\cite{CuHFS2}, the resonance linewidth is the combination of the laser bandwidth with the Doppler broadening from the hot atomizer temperature (typically 2500 K); in the case of in-gas-cell-laser spectroscopy, the Doppler contribution is limited to that of room temperature (300 K) but the pressure broadening plays an important role. A simulation of the respective contributions in case of the copper transition at 244.164 nm is shown in Fig.~\ref{fig:fig12}. Even at pressures as high as 500 mbar of Ar, the resolution for laser spectroscopy in a gas cell is better than with a hot cavity.

\begin{figure}
\begin{center}
 \includegraphics*[width=\columnwidth]{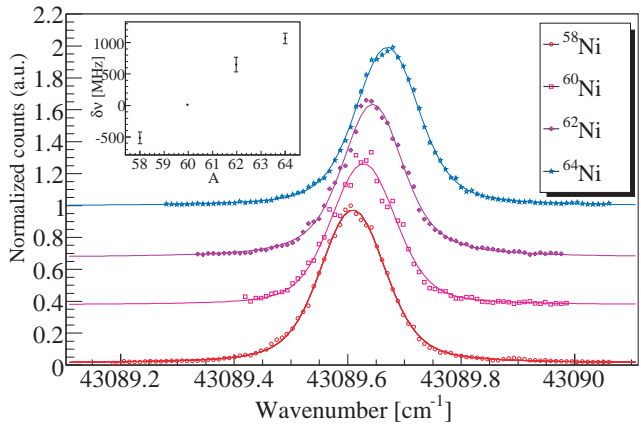}
\end{center}
\caption{Wavelength scan of the first step resonant transition in Ni at mass $A=$58,60,62,64 with He as the buffer gas in the LIST$_L$ (colour on-line). SPIG $V_{dc}$ $=20$ V, SPIG position was 0.75 mm from the gas cell. Insert: isotope shift of the even-A nickel isotopes.}
\label{fig13}
\end{figure}

The resolution of both systems remains however limited. This limit is mostly lifted when working in the LIST mode with the gas cell as the ions are cold and not under the influence of the pressure anymore. In Fig.~\ref{fig:fig12}, the present resolution with the LIST mode is given when the typical inherent laser bandwidth of 1.6 GHz is assumed. The resolution is dominated by the laser bandwidth. A laser with a narrower bandwidth could improve the resolution although the velocity distribution in the gas jet has to be taken into account (e.g.~Fig.~\ref{fig:fig10}). The reduction of the resonance linewidth in the LIST$_L$ opens the possibility of further laser spectroscopic studies at LISOL. This is demonstrated by measuring the isotope shift of $^{58,60,62,64}$Ni with either He or Ar as the buffer gas. Fig.~\ref{fig13} shows the result of the wavelength scans of the first step resonant transition for Ni at mass $A=58,60,62,64$ in He. Although the linewidth of the transition is still wide, mainly because of the laser bandwidth and power, the isotope shift of Ni was observed.

In order to relate the isotope shift to the changes in the mean-square charge radius, the electronic $F$-factor in the Field Shift (FS) and the Specific Mass Shift (SMS) have to be known. When isotope shifts have been measured for the same isotopes using different transitions, it is possible to extract a relative measurement of those parameters using a King plot. However, the uncertainty in our measurement is large in comparison with the limited contribution to the isotope shift of the nuclear effects. The linear relation yielding the relative information can therefore not be extracted. The extraction of the changes in the mean-square charge radius is thus impossible with our current setup in this mass range.

The Normal Mass Shift (NMS), one of the last contributions to the isotope shift, is indeed $\delta \nu \approx350$ MHz per two mass units, see the insert in Fig.~\ref{fig13}. This effect dominates the isotope shift and the FS, related to the changes in the mean-square charge radius, is buried underneath. The NMS and SMS become rapidly smaller as A increases while the FS increases with increasing Z; laser spectroscopy can therefore still be possible to determine the changes in the mean-square charge radius in heavier isotopic chains. Elements with large hyperfine parameters, like copper and bismuth, are good candidates for in-source laser spectroscopy to determine nuclear magnetic dipole moments \cite{CuHFS1}\cite{CuHFS2}\cite{BiHFS}.

\section{Conclusion}

The LIST coupled to a gas cell catcher has been studied at LISOL.
The operational novelty of this method is relying on element-selective resonant laser ionization of neutral atoms which is taking place inside the supersonic cold jet expanding out of the gas cell catcher. In this paper, some systematic studies have been performed with two different laser geometries, either longitudinal (LIST$_L$) or transverse (LIST$_T$) with respect to the gas jet outside the cell. In the LIST$_L$, a suppression voltage was applied on the SPIG $V_{dc}$. It follows that only photo-ions created inside the SPIG are sent to the mass separator; all other ions produced inside the cell are repelled. The needed suppression voltage was then found to be about 20V in He and 50V in Ar respectively. In the LIST$_T$, an ion collector voltage was utilized inside the gas cell. The suppression capability has been firstly demonstrated with neutron-deficient $^{94}$Rh isotopes produced in fusion-evaporation reactions. Although statistics was limited due to restrictions on the setup, this result shows that high selectivity is achievable.

Another aspect arising from the LIST is the feasibility for in-source laser spectroscopy after a gas cell. This possibility is opened by the extremely low density and low temperature inside the jet which makes the velocity distribution of the atoms nearly uniform, resulting in small pressure and Doppler broadenings. The resonance linewidth for the Ni isotopes at different locations was compared in terms of individual effects contributing into one FWHM. In these results, it can be found that the gas jet as an environment for laser spectroscopy is much more comparable to that of vacuum conditions, while inside the gas cell, the pressure is a crucial parameter for determining the resonance width. The broadening was evaluated to be 11.3 MHz/mbar for Ni and 5.4 MHz/mbar for Cu in Ar gas pressures between 60 and 530 mbar. In addition, the jet velocity for the two types of buffer gas has been evaluated as 1663 m/s for He and 550 m/s for Ar from the displacement of the resonance peak in the LIST$_L$. The possibility for isotope shift measurements with the stable even-A nickel isotopes was also demonstrated. For actual measurements of changes in the charge radius or magnetic moments of atomic nuclei, the FWHM of the resonance peak in the jet should be minimized for satisfying the demanded accuracy in either longitudinal or transverse approach. In the current setup, using a suitable laser system whose fundamental linewidth is typically 100 MHz, a resolution of the order of at best 1 GHz can be obtained due to the Doppler broadening inside the gas jet. However, a dedicated optimization of the gas jet should allow to further improve this resolution. Elements with large hyperfine parameters are still good candidates for ionization laser spectroscopy. Additionally this method will show a strong advantage for elements with slow release times or low release efficiencies in conventional ISOL systems as the gas flow transports all elements and decay losses can be minimized by fast evacuation of the cell volume. Spectroscopy inside the gas cell is also achievable for isotopes displaying large hyperfine structures or isotope shift; care should therefore be taken in choosing the appropriate conditions maximizing the production (higher pressures) while minimizing the resonance linewidth (lower pressures).

Certainly, for future applications of the LIST combined with a gas cell, the overlap efficiency between the laser photons and the jet atoms is a determining factor. The efficiency comprises two parameters: the time overlap and the geometrical overlap. For the enhancement of the former part, a high repetition laser system is needed to achieve at least one photon-atom encounter. Then the expected repetition rate depends on the ionization length which is related to the geometrical overlap. If the jet is collimated enough, for example, for 10 cm without expansion, the needed repetition rate is 5 kHz for one encounter (for a jet velocity of 500 m/s). Pulsed lasers satisfying such a repetition rate are now commercially available. For getting such a narrow jet, some special nozzle for the exit hole of the cell or specific pressure outside of the cell will be necessary \cite{blue}\cite{Kessler}. Finally, one should consider the losses associated to photo-ions produced in the gas cell and then repelled; such losses also appear in the vicinity of the exit hole when applying the suppression voltage between the SPIG and the cell exit hole. It can be minimized by sending the lasers from the other end of the beam line through the isotope separator and acceleration electrodes. Alternatively using the transverse approach with suppression voltage inside the cell is also suitable though a very high repetition laser is then needed.

\section{Acknowledgments}
We would like to thank the accelerator group at Louvain-La-Neuve for running and maintaining the accelerator. This work was supported by FWO-Vlaanderen (Belgium), GOA/2004/03 (BOF-K.U.Leuven), the 'Interuniversity Attraction Poles Programme - Belgian State - Belgian Science Policy'  (BriX network P6/23), and by the European Commission within the Sixth Framework Programme through I3-EURONS (Contract RII3-CT-2004-506065).


\begin{thebibliography}{00}
\bibitem{ls1} P.~Van Duppen, Nucl.~Inst.~and Meth.~B126(1997)66.
\bibitem{ls2} Yu.~Kudryavtsev, J.~Andrzejewski, N.~Bijnens, S.~Franchoo, J.~Gentens, M.~Huyse, A.~Piechaczek, J.~Szerypo, I.~Reusen, P.~Van~Duppen, P.~Van~den~Bergh, L.~Vermeeren, J.~Wauters, A. W\"ohr, Nucl.~Inst.~and Meth.~B114(1996)350. 
\bibitem{ls3} Yu.~Kudryavtsev, B.~Bruyneel, M.~Huyse, J.~Gentens, P.~Van~den~Bergh, P.~Van~Duppen, L.~Vermeeren, Nucl.~Inst.~and Meth.~B179(2001)412. 
\bibitem{ls4} Yu.~Kudryavtsev, B.~Bruyneel, S.~Franchoo, M.~Huyse, J.~Gentens, K.~Kruglov, W.F.~Mueller, N.V.S.V.~Prasad, R.~Raabe, I.~Reusen, P.~Van~den~Bergh, P.~Van~Duppen, J.~Van~Roosbroeck, L.~Vermeeren, L.~Weissman, Nucl.~Phys.~A701(2002)465.
\bibitem{ls5} M.~Huyse, M.~Facina, Yu.~Kudryavtsev, P.~Van~Duppen, ISOLDE collaboration, Nucl.~Inst.~and Meth.~B187(2002)535.
\bibitem{ls6} Yu.~Kudryavtsev, M.~Facina, M.~Huyse, J.~Gentens, P.~Van~den~Bergh, P.~Van~Duppen, Nucl.~Inst.~and Meth.~B204(2003)336. 
\bibitem{ls7} M.~Facina, B.~Bruyneel, S.~Dean, J.~Gentens, M.~Huyse,  Yu.~Kudryavtsev, P.~Van~den~Bergh, P.~Van~Duppen, Nucl.~Inst.~and Meth.~B226(2004)401. 
\bibitem{shd} Yu.~Kudryavtsev, T.E.~Cocolios, J.~Gentens, M.~Huyse, O.~Ivanov, D.~Pauwels, 
T.~Sonoda, P.~Van den Bergh, P.~Van Duppen, to be published.
\bibitem{list1} K.~Blaum, C.~Geppert, H.-J.~Kluge, M.~Mukherjee, S.~Schwarz and K.~Wendt, Nucl.~Inst.~and Meth.~B204(2003)331. 
\bibitem{list2} I.D.~Moore, A.~Nieminen, J.~Billowes, P.~Campbell, Ch.~Geppert, A.~Jokinen, T.~Kessler, B.~Marsh, H.~Penttila, S.~Rinta-Antila, B.~Tordoff, K.~Wendt and J.~\"Ayst\"o, J.~Phys.~G31(2005)s1499.
\bibitem{list3} P.~Karvonen, T.~Sonoda, I.D.~Moore, J.~Billowes, A.~Jokinen, T.~Kessler, H.~Penttila, A.~Popov, B.~Tordoff and J.~\"Ayst\"o, Eur.~Phys.~J.~Special Topics 150(2007)283. 
\bibitem{ig} J.~\"Ayst\"o, Nucl.~Phys~A693(2001)477.
\bibitem{sp1} H.J.~Xu, M.~Wada, J.~Tanaka, H.~Kawakami, S.~Ohtani and I.~Katayama, Nucl.~Inst.~and Meth.~A333(1993)274. 
\bibitem{sp2} S.~Fujitaka, M.~Wada, H.~Wang, J.~Tanaka, H.~Kawakami, I.~Katayama, K.~Ogino, H.~Katsuragawa, T.~Nakamura, K.~Okada, S.~Ohtani, Nucl.~Inst.~and Meth.~B126(1997)386.
\bibitem{sp3} P.~Van den Bergh, S.~Franchoo, J.~Gentens, M.~Huyse, Yu.~Kudryavtsev, A.~Piechaczek, R.~Raabe, I.~Reusen, P.~Van Duppen, L.~Vermeeren, A.~W\"ohr, Nucl.~Inst.~and Meth.~B126(1997)194.
\bibitem{sp4}P.~Karvonen, I.D.~Moore, T.~Sonoda, T.~Kessler, H.~Penttil\"a, K.~Per\"aj\"arvi, P.~Ronkanen, J.~\"Ayst\"o, Nucl.~Inst.~and Meth.~B266(2008)4794.
\bibitem{backe}H.~Backe, K.~Eberhardt, R.~Feldmann, M.~Hies, H.~Kunz, W.~Lauth, R.~Martin, H.~Schope, P.~Schwamb, M.~Sewtz, P.~Thorle, N.~Trautmann, S.~Zauner, Nucl.~Inst.~and Meth.~B126(1997)406. 
\bibitem{Yean}G.~Yeandle, J. Billowes, P. Campbell, P. Dendooven, K. Per\"aj\"arvi, M.D.~Seliverstov, G.~Tungate and J. \"Ayst\"o, Hyperfine Interactions 127(2000)91. 
\bibitem{strength}http://physics.nist.gov/PhysRefData/ASD/index.html.
\bibitem{Oli77}J.J.~Olivero and R.L.~Longbothum, J.~Quant.~Spectrosc.~Radiat.~Transfer,~Vol.~17,~pp.233-236 (1977).
\bibitem{Gatchina} G.D.
~Alkhazov, A.E.~Barzakh, V.P.~Denisov, K.A.~Mezilev, Yu.N.~Novikov, V.N.~Panteleyev, A.V.~Popov, E.P.~Sudentas, V.S.~Letokhov, V.I.~Mishin, V.N.~Fedoseyev, S.V.~Andreyev, D.S.~Vedeneyev, A.D.~Zyuzikov, Nucl.~Inst.~and Meth.~B69(1992)517.
\bibitem{CuHFS1} L.~Weissman, U.~K\"oster, R.~Catherall, S.~Franchoo, U.~Georg, O.~Jonsson, V.N.~Fedoseyev, V.I.~Mishin, M.D.~Seliverstov, J.~Van~Roosbroeck, S.~Gheysen, M.~Huyse, K.~Kruglov, G.~Neyens, P.~Van~Duppen, IS365~Collaboration~and~ISOLDE~Collaboration, Phys.~Rev.~C65(2002)024315.
\bibitem{PbCR} H.~De Witte, A.N.~Andreyev, N. Barr{\'e}, M.~Bender, T.E.~Cocolios, S.~Dean, D.~Fedorov, V.N.~Fedoseyev, L.M.~Fraile, S.~Franchoo, V.~Hellemans, P.H.~Heenen, K.~Heyde, G.~Huber, M.~Huyse, H.~Jeppessen, U.~K{\"o}ster, P.~Kunz, S.R.~Lesher, B.A.~Marsh, I.~Mukha, B.~Roussi{\`e}re, J.~Sauvage, M.~Seliverstov, I.~Stefanescu, E.~Tengborn, K.~Van de Vel, J.~Van de Walle, P.~Van Duppen, Yu.~Volkov, Phys.~Rev.~Letters 98(2007)112502.
\bibitem{CuHFS2} N.J.~Stone, U.~K\"oster, J.~Rikovska~Stone, D.V.~Fedorov, V.N.~Fedoseyev, K.T.~ Flanagan, M.~Hass, and S.~Lakshmi, Phys.~Rev.~C77(2008)067302.
\bibitem{BiHFS} I.D.~Moore, T.~Kessler, J.~\"Ayst\"o, J.~Billowes, P.~Campbell, B.~Cheal, B.~Tordoff, M.~L.~Bissel, G.~Tungate, Hyp.~Int.~171(2006)135.
\bibitem{blue} Bruce~Marsh, Doctor dissertation,~University~of~Manchester (2007).
\bibitem{Kessler} Thomas Kessler, Doctor dissertation,~University~of~Jyv{\"a}skyl{\"a} (2008).
\end{thebibliography}
\end{document}